\documentclass[conference]{IEEEtran}

\usepackage{eso-pic}

\newcommand{\ieeecopyrightnotice}{%
  \AddToShipoutPictureBG*{%
    \AtPageLowerLeft{%
      \put(\LenToUnit{\dimexpr(\paperwidth-\textwidth)/2\relax},%
           \LenToUnit{0.25in}){%
        \parbox{\textwidth}{\centering\scriptsize
          \copyright~2026 IEEE. Personal use of this material is permitted.
          Permission from IEEE must be obtained for all other uses, in any
          current or future media, including reprinting/republishing this
          material for advertising or promotional purposes, creating new
          collective works, for resale or redistribution to servers or lists,
          or reuse of any copyrighted component of this work in other works.
          Accepted for publication at IEEE QAI 2026.}%
      }%
    }%
  }%
}

\IEEEoverridecommandlockouts
\usepackage{cite}
\usepackage{amsmath,amssymb,amsfonts}
\usepackage{graphicx}
\usepackage{textcomp}
\usepackage{xcolor}
\usepackage{booktabs}
\usepackage[hidelinks]{hyperref}
\def\BibTeX{{\rm B\kern-.05em{\sc i\kern-.025em b}\kern-.08em
    T\kern-.1667em\lower.7ex\hbox{E}\kern-.125emX}}

\begin{document}


\title{FGSM Attack Transferability from Simulation to Quantum Hardware in Network Intrusion Detection}

\author{
\IEEEauthorblockN{Zeeshan Ajmal\textsuperscript{1},
Kimmo Halunen\textsuperscript{1},
Marc Mau\ss{}ner\textsuperscript{2},
Volker Reers\textsuperscript{3},
Arif Ali Khan\textsuperscript{1}}

 \IEEEauthorblockA{\textsuperscript{1}Faculty of Information
 Technology and Electrical Engineering (ITEE)\\
 University of Oulu, Finland\\
 \{zeeshan.ajmal, kimmo.halunen, arif.khan\}@oulu.fi}
 \IEEEauthorblockA{\textsuperscript{2}infoteam Software AG,
 Germany\\
 marc.maussner@infoteam.de}
 \IEEEauthorblockA{\textsuperscript{3}Qseidon GmbH, Germany\\
 volker.reers@qseidon.de}}

\maketitle
\ieeecopyrightnotice

\pagestyle{empty}
\thispagestyle{empty}

\begin{abstract}

Quantum computers are now available through cloud platforms, and recent
work has proposed using them to detect network intrusions. It is not yet
clear how easily such systems can be attacked. We train a hybrid
variational quantum classifier (VQC) on the Friday DDoS subset of
CICIDS2017 and test it against two attacks: label flipping during
training and FGSM evasion at prediction time. We run the tests in
simulation and on IBM hardware, the Fez superconducting processor
(Heron r2). We find that two models with similar accuracy can differ
greatly in how easily they are attacked. A model that decides with
little confidence loses half its detections to a very small change in
the input, while a model trained to decide with more confidence needs a
much larger change before it fails. Attacks built in simulation against
the more confident model also work on real hardware, and the device
classifies every unaltered attack sample correctly. Simulated device
noise does not reduce accuracy, but it does reduce the model's
confidence and makes attacks easier, so accuracy alone does not show how
exposed a model is. Label flipping has no measurable effect until 15\%
of attack labels are changed. These results show that device noise
should not be treated as a defense and that reported weaknesses in these
classifiers depend partly on how they were trained.
\end{abstract}


\section{Introduction}
\label{sec:intro}

In 1982, Feynman argued that classical computers cannot efficiently simulate quantum mechanical systems and that a quantum computer would be fundamentally more capable of such tasks~\cite{feynman1982}.
That argument has since extended beyond physics simulation to machine learning and cybersecurity~\cite{biamonte2017}. Quantum computers could accelerate machine learning by processing high-dimensional data in superposition, potentially discovering patterns that classical computers struggle with. In cybersecurity, quantum computers are being explored for network intrusion detection, anomaly detection, and cryptanalysis~\cite{abreu2025, kalinin2023}. Quantum computers are now accessible through cloud platforms and researchers have begun training quantum machine learning models on real security tasks.

Variational quantum classifiers (VQCs) are among the most practical quantum machine learning approaches for current hardware~\cite{havlicek2019}. A VQC encodes classical data into quantum states, processes it through
a trainable circuit, and produces a prediction via measurement. Current quantum computers operate in the noisy intermediate-scale quantum (NISQ) era~\cite{Preskill_2018}, characterized by 50 to 1000 qubits with significant decoherence and gate errors. 

Several studies have proposed VQC-based intrusion detection systems
and reported accuracy comparable to classical methods on standard
datasets~\cite{abreu2025, kalinin2023, qmlids2023}. However,
classical machine learning systems are well known to be vulnerable
to adversarial attacks at both training time and inference
time~\cite{goodfellow2015, biggio2012}. Whether these vulnerabilities
carry over to hybrid VQC systems on real security data has not
been studied.

This gap matters for two reasons. First, the security of a hybrid VQC depends on how much an attacker needs to know. An attacker with access to model gradients can target the decision boundary directly, while one limited to blind perturbation of the encoded inputs cannot. The size of
that difference determines how much gradient secrecy is worth as a defense. It is also unclear how much the defender's own training choices affect this, since a classifier that decides close to its boundary is easier to push across it than one that does not.
Second, prior adversarial studies on quantum classifiers have used simulators~\cite{lu2020, catak2023, maussner2025}. Prior work has documented discrepancies between simulated and real quantum processors in variational algorithm optimization~\cite{wang2021}, showing that simulator
behavior can diverge significantly from real hardware. Whether attacks crafted in simulation remain effective on real quantum hardware is an open question with direct consequences for how these systems should be tested before deployment.

We train a VQC on the CICIDS2017 Friday DDoS subset~\cite{cicids2017}, apply FGSM (Fast Gradient Sign Method, a gradient-guided attack)~\cite{goodfellow2015}
evasion and label-flipping poisoning attacks, compare gradient-guided against untargeted perturbations and measure how the trained decision confidence affects vulnerability, and validate results on the IBM Fez
processor. Three research questions guide the study:

\begin{itemize}
  \item \textbf{RQ1:} At what rate does label-flipping poisoning
        degrade a VQC-based intrusion detector?
  \item \textbf{RQ2:} How much does gradient guidance increase
        evasion success over untargeted perturbation of the same
        inputs, and how does the trained decision confidence
        affect this?
  \item \textbf{RQ3:} Do adversarial examples crafted in simulation
        remain effective on real quantum hardware, and does
        simulated device noise change that exposure?
\end{itemize}

This paper contributes a simulation-to-hardware transfer measurement for
an intrusion detection VQC, evidence that decision margin rather than
accuracy predicts adversarial exposure, and a controlled comparison
showing that simulated device noise raises attack success while leaving
accuracy unchanged.


\section{Background}
\label{sec:background}

\subsection{Variational Quantum Classifiers}

 A quantum computer operates on qubits, which can exist in
superposition of $|0\rangle$ and $|1\rangle$ and become entangled
with one another~\cite{biamonte2017}. Variational quantum classifiers
(VQCs) use these properties for classification tasks. A VQC has three components. First, a data encoding layer maps classical 
input features to qubit rotation angles, where each input feature is 
converted into an angle that rotates a qubit. Second, a variational ansatz 
(a trainable quantum circuit) applies a sequence of rotation gates and 
entangling gates to process the encoded quantum state. Third, a measurement 
step extracts a classical prediction by measuring a Pauli operator, which is 
a quantum observable that represents a measurement direction in quantum space. 

The magnitude of the measured expectation value indicates how far a sample
sits from the decision boundary. A value near zero means the classifier is
close to changing its prediction, while a value near one means the
prediction is stable under small changes to the input. We refer to the
median of this magnitude over correctly classified samples as the decision
margin, and use it throughout to characterize trained models.

\subsection{Adversarial Attacks on ML and QML}

Adversarial attacks on machine learning models were first studied
systematically by Szegedy et al.~\cite{szegedy2014}, who showed that
small input perturbations can cause deep networks to misclassify with
high confidence. Goodfellow et al.~\cite{goodfellow2015} introduced the
Fast Gradient Sign Method (FGSM). Biggio et al.~\cite{biggio2012} showed
that corrupting a small fraction of training labels is sufficient to
degrade classifier performance at test time.

These attacks have been extended to quantum classifiers. Lu
et al.~\cite{lu2020} showed that quantum classifiers are vulnerable
to adversarial perturbations and that classical-side attacks transfer
to the quantum setting. Catak, Seo, and Cali~\cite{catak2023} evaluated attacks on classical inputs and quantum
states in a two-qubit variational quantum classifier trained on synthetic
data. Gradient-guided FGSM and PGD (Projected Gradient Descent), an iterative 
variant of FGSM that applies multiple gradient steps with projection to maintain 
perturbation constraints~\cite{madry2018}, applied to the encoded inputs achieved
substantially higher attack success rates than untargeted random
perturbations of the same inputs, indicating that access to gradient
information is a significant advantage for an attacker. Mau{\ss}ner and
Reers~\cite{maussner2025} extended this to noisy simulators, testing
FGSM and PGD attacks and evaluating adversarial retraining as a
defense. Kundu and Ghosh~\cite{kundu2022} introduced QUID, a
poisoning attack that targets the quantum encoding structure directly.
Ghosh et al.~\cite{ghosh2023} survey adversarial threats across the QML pipeline and identify vulnerabilities at the data encoding, circuit, and deployment stages.

\subsection{QML for Network Intrusion Detection}

Several studies have applied QML to network intrusion detection.
Abreu et al.~\cite{abreu2025} proposed QuantumNetSec, a hybrid
VQC-based system that achieves competitive accuracy on CICIDS2017~\cite{cicids2017}. Kalinin and Krundyshev~\cite{kalinin2023} evaluated quantum machine learning
for intrusion detection using a custom stream dataset derived from
IoT network traffic~\cite{kang2019iot}. Abreu et al.~\cite{qmlids2023} demonstrated binary DDoS classification
using a VQC trained on CICIDS2017.

Table~\ref{tab:related} compares these studies against this paper on
four criteria. The intrusion detection studies report accuracy, and two
of them run on real IBM hardware, but none tests adversarial attacks.
The adversarial studies test attacks, but on simulators and on tasks
other than intrusion detection. No prior work combines an adversarial
evaluation with hardware validation, and none reports a decision margin.

\begin{table}[t]
\caption{Coverage of related work compared with this paper. IDS:
intrusion detection task. Adv.: adversarial attacks tested. HW: real
quantum hardware. Margin: decision margin reported.}
\label{tab:related}
\centering
\begin{tabular}{lcccc}
\toprule
\textbf{Study} & \textbf{IDS} & \textbf{Adv.} & \textbf{HW} & \textbf{Margin} \\
\midrule
Abreu et al.~\cite{abreu2025}          & \checkmark & & \checkmark & \\
Abreu et al.~\cite{qmlids2023}         & \checkmark & & \checkmark & \\
Kalinin and Krundyshev~\cite{kalinin2023}      & \checkmark & & & \\
Lu et al.~\cite{lu2020}                & & \checkmark & & \\
Catak et al.~\cite{catak2023}          & & \checkmark & & \\
Mau{\ss}ner and Reers~\cite{maussner2025} & & \checkmark & & \\
\midrule
This paper & \checkmark & \checkmark & \checkmark & \checkmark \\
\bottomrule
\end{tabular}
\end{table}

\section{Methodology}
\label{sec:methodology}

\subsection{Dataset and Preprocessing}

We use the CICIDS2017 dataset~\cite{cicids2017}, specifically the
Friday afternoon DDoS traffic subset, which provides a binary
classification task (benign vs.\ DDoS). The dataset contains 78
statistical flow features extracted from raw packet captures.
Preprocessing follows four steps: (1) rows with infinite or missing
values are removed; (2) all features are scaled to $[0,1]$ using
min-max normalization; (3) principal component analysis (PCA) reduces
the feature space from 78 to 4 dimensions, corresponding to the number
of qubits, retaining approximately 62\% of total variance; (4) the PCA
output is rescaled to $[0,\pi]$ for compatibility with angle encoding.
The reduction to four components is fixed by the qubit count rather than
chosen for variance retention. The 92.2\% clean accuracy reported in
Section~\ref{sec:results} indicates that the retained components separate
the two classes adequately for this binary task, though a wider circuit
would allow more of the feature space to be preserved. The scaling and
PCA transforms are fitted on the full dataset before the train and test
split. Both are unsupervised and use no label information, but the test
features do influence the fitted transforms; we return to this in
Section~\ref{sec:threats}.

We use 2000 training samples and 500 test samples, selected with
stratified sampling to preserve class balance. Of the 2000 training
samples, 400 are held out as a stratified validation set used for model
selection, leaving 1600 for training. The test set is used only for the
final reported measurement. The test subset contains 284 attack and 216
benign samples.

\subsection{VQC Architecture}
\label{sec:vqc}

The VQC uses 4 qubits and 3 variational layers, giving 24 trainable
parameters, implemented in PennyLane~\cite{pennylane}. Each qubit $i$ is initialized by a rotation $R_Y(x_i)$, where 
$R_Y(\theta) = \exp(-i\theta Y/2)$ is a single-qubit rotation 
about the $Y$-axis by angle $\theta$, and $x_i$ is the $i$-th 
preprocessed input feature. Each variational layer applies 
$R_Y(\theta_{l,i,0})$ and $R_Z(\theta_{l,i,1})$ to each qubit, 
where $R_Z(\theta) = \exp(-i\theta Z/2)$ is a rotation about the 
$Z$-axis, followed by CNOT entangling gates in a ring topology. The circuit
output is the Pauli-$Z$ expectation value on qubit 0, mapped to a
binary prediction: $+1$ for benign traffic and $-1$ for attack
traffic.

Training minimizes a hinge loss over the full training set using the Adam
optimizer~\cite{kingma2017} with a learning rate of 0.05 for 150 epochs.
Gradients are computed analytically by backpropagation through the
simulated circuit, which is tractable at this circuit size. Five
independent initializations are trained per configuration, each drawn
from a seeded generator so that runs are reproducible. The reported model
is the one with the highest validation accuracy; the test set is used
only for the final measurement. We also report the mean and standard
deviation of test accuracy across initializations, since variational
circuits of this size are sensitive to where optimization begins.

Two models are trained for the evasion experiments. The first follows the procedure above. The second uses the gradient-free COBYLA optimizer~\cite{powell1994} with a budget of 500 iterations, minimizing the same hinge loss over random mini-batches of 32 samples. The two models share architecture, data, and preprocessing. They differ in optimizer and in how the loss is estimated. COBYLA converges early on the
noisy mini-batch objective, so the second model reaches similar clean accuracy but places its correctly classified samples much closer to the
decision boundary. The second model is included to give a pair that differs in margin at comparable accuracy, which is what the evasion comparison in Section~\ref{sec:rq2} requires. Optimizer and margin are confounded in this pair. Section~\ref{sec:threats} reports a control that
did not separate them. The two are referred to below as the wide-margin and narrow-margin models.

\subsection{Attack Design}
Three attacks are used. The first changes labels in the training set. The other two change feature values at test time, after PCA and scaling, so a
perturbation moves the encoded rotation angles rather than the raw flow features.

\subsubsection{Label-Flipping Poisoning (RQ1)}
A fraction of attack-labeled samples is relabeled as benign before
training. Flipping is applied to the full 2000-sample pool before the
validation split, so the validation labels carry the same corruption and
the defender cannot select a model on clean data. This simulates a
malicious data supplier who corrupts the training set without knowledge
of the model architecture or parameters. We test poison rates of 5\%, 10\%,
and 15\% of attack-labeled samples. Attack success rate (ASR) is
defined as the fraction of true attack samples in the test set that
the poisoned model misclassifies as benign, measured over all attack
samples rather than only those classified correctly as in the evasion
case below.

\subsubsection{FGSM Evasion (RQ2)}
At inference time, FGSM~\cite{goodfellow2015} crafts adversarial
examples using:
\begin{equation}
\mathbf{x}_{adv} = \text{clip}\!\left(\mathbf{x} +
\varepsilon \cdot \text{sign}\!\left(\nabla_\mathbf{x}
\mathcal{L}(f(\mathbf{x}), y)\right),\, 0,\, \pi\right)
\end{equation}
where $\varepsilon \in \{0.05, 0.10, 0.15, 0.20, 0.30, 0.40, 0.50,
0.75, 1.00\}$ is the perturbation budget, $\mathcal{L}$ is the hinge
loss, and $\text{clip}$ keeps each element within the valid input domain
$[0, \pi]$. We also report $\varepsilon$ as a percentage of this range,
so $\varepsilon = 0.30$ is a perturbation of 9.5\%. Gradients are computed with respect to the
classical input features using PennyLane's automatic differentiation
interface. ASR is measured only over attack samples that the model
classifies correctly before perturbation, so that an attack counts as
successful only when it changes a decision the model would otherwise
have made correctly. More powerful iterative variants such as
PGD~\cite{madry2018} are left for future work.

\subsubsection{Untargeted Perturbation Baseline (RQ2)}
As a baseline, uniform random noise within the same budget is applied
to the encoded input angles without gradient guidance. This simulates
an attacker who can reach the encoding layer but has no access to
gradient information. Because the encoding applies $R_Y(x_i)$
directly, the rotation angles and the preprocessed input features are
the same quantities, so this baseline perturbs exactly what FGSM
perturbs. The two therefore differ only in whether the perturbation is
guided by the gradient, and comparing them isolates the value of
gradient access to an attacker rather than the relative exposure of
the classical and quantum components.

\subsection{Noise Modeling (RQ3)}
To approximate NISQ hardware conditions, a depolarizing channel is
applied after every gate:
\begin{equation}
\mathcal{E}(\rho) = (1-p)\rho + \frac{p}{3}(X\rho X + Y\rho Y + Z\rho Z)
\end{equation}
at noise levels $p \in \{0.000, 0.001, 0.010, 0.050\}$, corresponding to
noiseless, low-noise, moderate-noise, and high-noise configurations
respectively. The channel is applied after each gate, so $p$ is a
per-gate error rate. On the IBM Fez backend used for the hardware experiments, the median
one-qubit gate error is of order $10^{-4}$ and the median two-qubit gate
error is of order $10^{-3}$. The low level falls between
these two rates, the moderate level is a few times the two-qubit rate,
and the high level is well above anything the device reports. The
simulated channel is uniform across gates, so these levels bracket the
measured rates rather than reproduce them. 

Density matrix simulation is substantially more costly than statevector
simulation at this circuit size, so the noisy configurations are trained
on a fixed subsample of 400 training samples with three initializations
per level. The subsample is fixed rather than resampled, so the training
objective remains deterministic. For comparison across noise levels we
also train the noiseless configuration on the same 400 samples, since
accuracy at different training set sizes is not directly comparable. The
full-size noiseless result is reported separately.

\subsection{IBM Quantum Hardware Validation (RQ3)}
We ran experiments on the IBM Fez superconducting processor (Heron r2, 156
qubits) at 1024 shots per circuit via Qiskit Runtime SamplerV2. Two jobs
were submitted, each as a single batch so that all its circuits share one
calibration state. The first evaluates the clean and 10\% poisoned models
on the same stratified subset of 50 test samples; alongside accuracy we
report the fraction of samples on which hardware and simulator return the
same label, which separates device error from model accuracy. Simulator
values reported alongside hardware results are computed on these same
50-sample subsets, so they differ slightly from the full test set values
in Section~\ref{sec:rq2}. The second
evaluates adversarial transfer: for 50 attack samples the clean model
classifies correctly in simulation, we submit the unperturbed input and
its FGSM counterpart at $\varepsilon \in \{0.30, 0.50\}$. Adversarial
examples are crafted against the simulator and submitted without
hardware-specific modification. An attack counts as successful only if
hardware classifies the unperturbed sample correctly and the perturbed
sample incorrectly. Circuits were transpiled at optimization level 1,
raising depth from 20 to between 87 and 100 and two-qubit gate count from
12 to between 27 and 42 depending on the qubit mapping.

\subsection{Evaluation Metrics}
We report clean accuracy (fraction of test samples correctly classified),
attack success rate (ASR, defined separately for each attack above), false
positive rate (FPR, fraction of benign samples misclassified as attack),
and the decision margin defined in Section~\ref{sec:background}.
Proportions in Tables~\ref{tab:poisoning}, \ref{tab:hardware} and
\ref{tab:noise} carry a 95\% Wilson score interval. Where a quantity
varies across random initializations or random draws, we report the mean
and standard deviation in addition to the selected model.

\subsection{Implementation Provenance}
FGSM follows Goodfellow et al.~\cite{goodfellow2015}. Circuit templates,
optimizers and automatic differentiation are standard PennyLane
components~\cite{pennylane}; hardware execution uses Qiskit Runtime.
Preprocessing follows common practice for CICIDS2017, and label-flipping
poisoning follows the standard formulation~\cite{biggio2012}. The
simulate-then-execute transfer protocol, the margin-controlled model
comparison, the untargeted perturbation baseline as constructed here, and
the evaluation pipeline connecting them are contributions of this work.
All code and result files are available
online.\footnote{\url{https://github.com/xeeshanajmal/qai2026-fgsm-transfer}}

\section{Results}
\label{sec:results}

\subsection{Clean Baseline}
The classifier reaches 85.5\% mean test accuracy across five random
initializations (standard deviation 8.0) when trained on the full
1600-sample split, with the validation-selected model reaching 92.2\%.
The spread is bimodal rather than continuous, which is why we report it
across initializations throughout; Section~\ref{sec:threats} gives the
detail.

The margin-controlled comparison uses two models of similar clean
accuracy but different decision margin: the wide-margin model above, with
a median $|\langle Z_0 \rangle|$ of 0.54 over correctly classified attack
samples, and a narrow-margin model reaching 86.8\% accuracy with a median
of 0.07. The narrow-margin accuracy lies within the range spanned by
individual initializations of the wide-margin model, so the two are not
separated by accuracy.

\subsection{RQ1: Poisoning Attack Results}
Table~\ref{tab:poisoning} shows label-flipping results across three
poison rates.

\begin{table}[t]
\caption{Label-flipping poisoning results.}
\label{tab:poisoning}
\centering
\begin{tabular}{lccc}
\toprule
\textbf{Poison Rate} & \textbf{Clean Acc.} & \textbf{ASR} & \textbf{FPR} \\
\midrule
5\%  & 86.8\% $\pm$ 8.3 & 0.0\% [0.0, 1.3]   & 11.1\% [7.6, 15.9] \\
10\% & 90.2\% $\pm$ 8.8 & 0.0\% [0.0, 1.3]   & 5.1\% [2.9, 8.9] \\
15\% & 76.4\% $\pm$ 0.8 & 38.7\% [33.3, 44.5] & 2.8\% [1.3, 6.0] \\
\bottomrule
\end{tabular}
\end{table}

At 5\% and 10\%, flipping 56 and 113 of the 1134 attack labels in the
training pool produces no measurable attack success: every true attack in the test set is still
classified as an attack. Clean accuracy at these rates is not
distinguishable from the unpoisoned baseline, nor from each other, given
the spread across initializations.

At 15\%, where 170 labels are flipped, the attack succeeds on 38.7\% of attack samples. The transition
is abrupt rather than gradual. It is also the most consistent result in
our experiments: all five initializations converge to within 0.8
percentage points of the same accuracy, compared with spreads above eight
points at the lower rates. Rather than degrading the classifier by
degrees, corruption beyond some point between 10\% and 15\% drives every
run to the same degraded solution.

Falling FPR across the three rates is consistent with this reading. As
more attack samples are labeled benign, the model shifts toward
predicting benign, which reduces false alarms while allowing genuine
attacks through.

\subsection{RQ2: Gradient Guidance and Decision Margin}
\label{sec:rq2}

Table~\ref{tab:evasion} compares FGSM against untargeted perturbation of
the same inputs at four representative budgets from the sweep. 

\begin{table}[t]
\caption{FGSM against untargeted perturbation. Untargeted results are
averaged over ten random draws.}
\label{tab:evasion}
\centering
\begin{tabular}{llccc}
\toprule
\textbf{Model} & $\varepsilon$ & \textbf{FGSM} & \textbf{Untargeted} & \textbf{Ratio} \\
\midrule
Narrow & 0.05 & 48.9\% & 2.0\% $\pm$ 1.8 & 24.4$\times$ \\
margin & 0.10 & 53.3\% & 11.6\% $\pm$ 4.5 & 4.6$\times$ \\
$|\langle Z_0\rangle| = 0.07$ & 0.20 & 60.0\% & 17.3\% $\pm$ 7.1 & 3.5$\times$ \\
 & 0.50 & 91.1\% & 26.7\% $\pm$ 7.1 & 3.4$\times$ \\
\midrule
Wide & 0.05 & 0.0\% & 0.0\% & n/a \\
margin & 0.10 & 0.0\% & 0.0\% & n/a \\
$|\langle Z_0\rangle| = 0.54$ & 0.20 & 5.0\% & 0.0\% & n/a \\
 & 0.50 & 71.7\% & 3.5\% $\pm$ 1.4 & 20.5$\times$ \\
\bottomrule
\end{tabular}
\end{table}

Gradient guidance is worth a large factor to an attacker, and the factor
is largest where the attacker is most constrained. On the narrow-margin
model, FGSM at $\varepsilon = 0.05$ succeeds 24 times more often than
untargeted noise of the same budget; by $\varepsilon = 0.50$ the
advantage falls to roughly three times, as random perturbation of
sufficient size begins to cross the boundary on its own. Because both
attacks perturb the same rotation angles, this measures the value of
gradient access rather than any asymmetry between the classical and
quantum parts of the model.

Decision margin has the larger effect. The narrow-margin model loses
48.9\% of its detections at $\varepsilon = 0.05$, a perturbation of 1.6\%
of the feature range. The wide-margin model is unaffected at that budget
and at 0.10, reaches 5.0\% at 0.20, and first approaches comparable
attack success near $\varepsilon = 0.50$, ten times larger. The two
models classify clean traffic to within a few percentage points of each
other. The narrow-margin model also misclassifies 15 of 60 attack samples
before any attack is applied, while the wide-margin model classifies all
60 correctly.

A classifier can therefore appear adequate by accuracy alone while
deciding so close to its boundary that a minimal perturbation moves it
across.

\subsection{RQ3: Simulation-to-Hardware Transferability}
\label{sec:rq3-transfer}

\begin{figure}[t]
\centerline{\includegraphics[width=0.78\columnwidth]{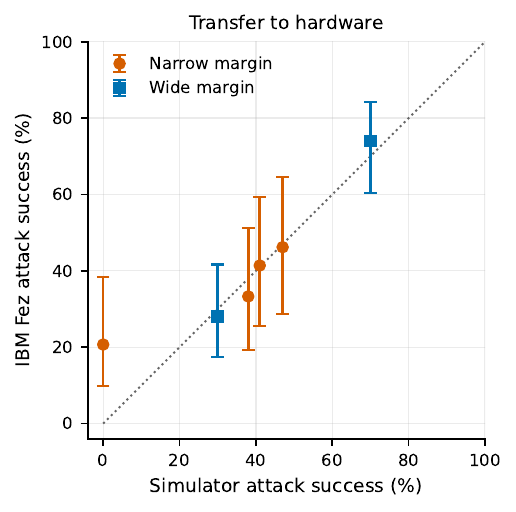}}
\caption{Attack success on IBM Fez against the value predicted in
simulation, six budgets across both classifiers. The dotted line marks
equality. At the leftmost point the perturbation is negligible, so the
measured successes are device error.}
\label{fig:transfer}
\end{figure}

Table~\ref{tab:hardware} and Fig.~\ref{fig:transfer} show results measured
on IBM Fez. The narrow-margin model was evaluated on hardware at four
further budgets, giving six paired simulator and hardware measurements in
total.

\begin{table}[t]
\caption{Simulation against IBM Fez hardware. Intervals are 95\% Wilson.}
\label{tab:hardware}
\centering
\begin{tabular}{lcc}
\toprule
\textbf{Measurement} & \textbf{Simulator} & \textbf{Hardware} \\
\midrule
Clean model accuracy & 90.0\% & 92.0\% [81, 97] \\
Poisoned model accuracy & 96.0\% & 92.0\% [81, 97] \\
\midrule
FGSM ASR, $\varepsilon = 0.30$ & 30.0\% & 28.0\% [17, 42] \\
FGSM ASR, $\varepsilon = 0.50$ & 70.0\% & 74.0\% [60, 84] \\
\bottomrule
\end{tabular}
\end{table}

The device reproduces the simulated classifier closely. Hardware and
simulator return the same label on 94\% of samples for the clean model
and 92\% for the poisoned model, so the small accuracy differences in
Table~\ref{tab:hardware} reflect device error on a minority of samples
rather than a systematic shift in the decision boundary.

Adversarial examples transfer without measurable loss. At both budgets in
Table~\ref{tab:hardware} the simulated attack success rate falls inside
the hardware confidence interval, and in that experiment the hardware
classified all 50 unperturbed attack samples correctly, so every observed
failure is attributable to the perturbation rather than to device error. This holds across a transpilation that
raises two-qubit gate count from 12 to between 27 and 42 depending on the
qubit mapping, and at two budgets differing by more than a factor of two
in attack success.

\subsection{RQ3: Effects of Simulated Device Noise}
\label{sec:rq3-noise}

\begin{figure}[t]
\centerline{\includegraphics[width=\columnwidth]{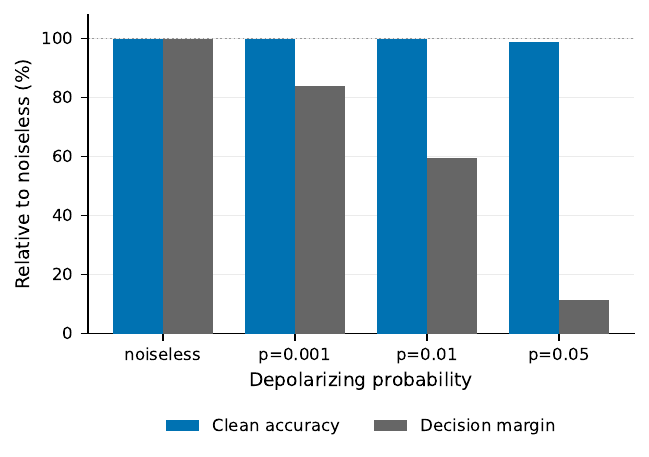}}
\caption{Clean accuracy and decision margin under depolarizing noise,
relative to their noiseless values.}
\label{fig:noise}
\end{figure}

Table~\ref{tab:noise} and Fig.~\ref{fig:noise} show the effect of
depolarizing noise on clean accuracy, decision margin, and attack
success. All configurations in this comparison are trained on the same
400-sample subset, so accuracy is comparable across noise levels.

\begin{table}[t]
\caption{Effect of depolarizing noise. Accuracy is the mean over three
initializations; FGSM is measured at $\varepsilon = 0.30$.}
\label{tab:noise}
\centering
\begin{tabular}{lccc}
\toprule
\textbf{Noise} & \textbf{Clean Acc.} & \textbf{Margin} & \textbf{FGSM ASR} \\
\midrule
Noiseless   & 89.2\% $\pm$ 7.1  & 0.54 & 33.3\% [23, 46] \\
$p = 0.001$ & 89.1\% $\pm$ 7.2  & 0.46 & 36.7\% [26, 49] \\
$p = 0.01$  & 88.9\% $\pm$ 7.1  & 0.32 & 36.7\% [26, 49] \\
$p = 0.05$  & 88.2\% $\pm$ 10.6 & 0.06 & 53.3\% [41, 65] \\
\bottomrule
\end{tabular}
\end{table}

Clean accuracy is unchanged across two orders of magnitude of
depolarizing probability. The largest difference between any two levels
is one percentage point, far smaller than the spread across random
initializations at any single level. By this measure alone, noise at
these rates has no effect on the classifier.

The decision margin tells a different story. It falls steadily from 0.54
to 0.06 as noise increases, reaching 11\% of its noiseless value while
accuracy holds at 99\% of its own (Fig.~\ref{fig:noise}), and attack
success rises from 33.3\% to 53.3\% at a fixed perturbation budget. A depolarizing channel contracts
Pauli expectation values by construction, so part of this reduction is
mechanical rather than learned. Contraction alone preserves the sign of
the expectation value and therefore cannot change a prediction, so it
does not account for the increase in attack success; that increase
reflects a genuine reduction in how far samples sit from the boundary.

Accuracy alone is therefore not sufficient to assess a variational
classifier for deployment. A model may hold its accuracy under noise
while becoming markedly easier to attack.


\section{Discussion}
\label{sec:discussion}

\subsection{Decision Margin as a Security Property}
The two models in Section~\ref{sec:rq2} differ in clean accuracy by a few
percentage points and in adversarial exposure by roughly a factor of ten
in required perturbation budget. Accuracy does not predict this
difference; margin does. The same relationship appears again under noise,
where accuracy is flat across two orders of magnitude while margin falls
by a factor of nine and attack success rises accordingly.

Margin is therefore worth measuring and reporting when variational
classifiers are evaluated for security applications. It costs nothing
beyond the predictions already computed, and it separates models that
accuracy alone treats as equivalent. It also offers one reading of the
wide range of adversarial vulnerability reported for variational
classifiers~\cite{lu2020, catak2023, maussner2025}: results obtained from
models trained to different margins are not directly comparable, even at
matched accuracy.

\subsection{Simulation is Sufficient for the Attacker}

Adversarial examples crafted against a noiseless simulator succeeded on
IBM Fez at the rate the simulator predicted, at both budgets tested, with
no hardware-specific modification. The device also classified every
unperturbed attack sample correctly, so device error did not contribute
to the measured attack success.

An attacker needs no access to the target device. Knowledge of the model
is sufficient, and the noise and transpilation overhead of current
hardware do not degrade a transferred attack. Device noise should not be
treated as an incidental defense.

For evaluation this is a convenience: adversarial testing of models at this scale can be conducted in simulation and is likely to reflect deployed behavior. 

Reers and Mau{\ss}ner~\cite{marc2026} describe a deployment shift effect,
where adversarial examples tuned to one device noise profile lose
effectiveness under different calibration data or coupling maps. Our
attacks are crafted against a noiseless simulator and are therefore not
fitted to any particular profile, so this effect does not arise here. At
4 qubits and 3 layers the device tracks the simulator closely enough that
untuned attacks transfer without loss. 

\subsection{Implications for VQC-Based IDS Deployment}
Three observations bear on deployment. First, label-flipping poisoning
showed no effect below a threshold and a sharp degradation above it.
Monitoring clean accuracy would not reveal corruption at 5\% or 10\%,
because there is nothing to reveal, but it would reveal corruption at
15\%. Data provenance matters, though within the range tested the risk is
concentrated rather than gradual.

Second, gradient access is the largest single advantage available to an
attacker at small perturbation budgets, where it was worth a factor of
24. Model confidentiality is therefore a meaningful part of a defense,
though it becomes less valuable as the budget grows.

Third, a model that holds its accuracy under noise may nonetheless be
substantially easier to attack, and acceptance testing that checks
accuracy alone will not detect this. Because attack success rose
monotonically with depolarizing probability in our experiments, an
adversary able to influence device noise, through scheduling, crosstalk
from co-located circuits, or calibration timing, would increase attack
effectiveness by doing so. This threat vector has been noted in the
broader QML security literature~\cite{ghosh2023}; our results give it a
measurable direction, at least under a depolarizing model.

\section{Threats to Validity}
\label{sec:threats}

\textit{Internal validity.} Variational circuits at this size are
sensitive to initialization. Test accuracy across five initializations
falls into two groups, near 92\% and near 75\%, with none in between. We
report mean and standard deviation across initializations throughout, and
select on a held-out validation split rather than on test accuracy, but a
spread of this size means single-run results at this scale should be
treated with caution. Scaling and PCA are fitted on the full dataset
before the train and test split. Both are unsupervised and use no label
information, but the test features influence the fitted transforms, which
may make absolute accuracies slightly optimistic. The comparisons in this
paper are between models sharing the same transforms, so this does not
affect the relationships we report.

\textit{Construct validity.} The untargeted baseline perturbs the same
rotation angles as FGSM and differs only in the absence of gradient
guidance. It therefore measures the value of gradient access, not the
relative exposure of the classical and quantum components of the model;
separating those would require a gradient-guided attack constructed in
the quantum state space. Attack success rate is
measured only over samples classified correctly before perturbation,
which makes it comparable across models of differing accuracy. Decision
margin is measured as the median $|\langle Z_0\rangle|$ over correctly
classified attack samples; under a depolarizing channel this quantity is
contracted mechanically as well as by any change in the learned
boundary, so margin values across noise levels should not be read as a
pure measure of learned confidence. Varying training length within a
single optimizer did not separate margin from optimizer choice. Across
18 runs, six checkpoints for each of three initializations, median margin ranged only
from 0.22 to 0.75 and never approached the 0.07 of the narrow-margin
model, so this control did not reach the regime where the effect was
observed. Within the range it did cover the relationship was
inconsistent, and margin was confounded with convergence, since the runs
reaching high accuracy were also the ones that became attackable. Margin
is therefore associated with exposure in our data rather than shown to
drive it.

\textit{External validity.} The results come from one dataset, one binary classification task, one small circuit, and one hardware device.
They may not hold for multi-class intrusion detection, other attack types, larger circuits, or other architectures. The hardware findings
were obtained at a depth where the device tracks the simulator for over 90\% of samples. At greater depth, where more noise accumulates, transfer behavior may differ. Establishing whether these findings hold more widely would need the same comparison repeated on other datasets, on multi-class tasks, on larger circuits, and on more than
one device.

\textit{Conclusion validity.} Hardware results are based on 50 samples
per condition, and the resulting Wilson intervals are correspondingly
wide. The transfer conclusion rests on the simulated rate falling inside
those intervals at both budgets tested rather than on a narrow estimate
at either. Noise experiments use three initializations rather than five
and a reduced training subsample, owing to the cost of density matrix
simulation. Agreement between simulator and hardware on over 90\% of
samples supports the use of simulation at this circuit scale, but
depolarizing noise is a simplified model that omits readout error and
crosstalk~\cite{wang2021, ghosh2023}. Device-calibrated noise models such as FakeFez would improve 
simulator fidelity and remain future work.

\section{Conclusion}
\label{sec:conclusion}

Adversarial exposure depends on more than accuracy. Two models of the
same architecture and similar clean accuracy differed by roughly a factor
of ten in the perturbation budget needed to defeat them, and decision
margin is the measurable property that separates them. The same pattern
appears under simulated noise, where accuracy is unchanged across two
orders of magnitude while margin falls and attack success rises.

Adversarial examples crafted in simulation transfer to IBM Fez without
measurable loss at both budgets tested, and the device classified every
unperturbed attack sample correctly. An attacker needs no knowledge of
the target device, and device noise offers no incidental protection.
Label-flipping poisoning had no measurable effect below 15\% of attack
labels, above which every initialization converged to the same degraded
solution.

Together these results suggest that security evaluation of variational
classifiers should report decision margin alongside accuracy, and that
simulation is an adequate setting for adversarial testing at this circuit
scale. 

Future work includes iterative attacks such as PGD, adversarial
retraining, gradient-guided perturbation in the quantum state space, PCA
ablation studies, optimizer comparisons, reduced-depth circuit
designs~\cite{West_2024}, larger and deeper circuits where accumulated
noise may change transfer behavior, and robustness under device recalibration 
and changing coupling maps~\cite{maussner2025, marc2026}.

\section*{Acknowledgment}

This work was supported by the Business Finland project (24304955) SeQuSoS (Secure Quantum Software Systems). The authors acknowledge the SeQuSoS research group for providing a collaborative research environment.

We acknowledge the use of IBM Quantum services for this work~\cite{ibmquantum}. The views expressed are those of the authors and do not reflect 
the official policy or position of IBM or the IBM Quantum team. In this work we used \texttt{ibm\_fez}, which is one of the IBM Quantum Heron r2 processors. Claude (Anthropic, 2025)~\cite{claude2025} was used to assist with developing the experimental code and with drafting and revising the manuscript text. Grammarly was used for language refinement. The authors designed the study, conducted all experiments, verified every reported result against the stored output, and take full responsibility for the content of this paper.

\bibliographystyle{IEEEtran}
\bibliography{references}

@article{feynman1982,
  title = {Simulating Physics with Computers},
  author = {Feynman, Richard P.},
  year = {1982},
  date = {1982-06-01},
  journal = {International Journal of Theoretical Physics},
  shortjournal = {International Journal of Theoretical Physics},
  volume = {21},
  number = {6},
  pages = {467--488},
  issn = {1572-9575},
  doi = {10.1007/BF02650179},
  url = {https://doi.org/10.1007/BF02650179}
}

@article{West_2024,
   title={Drastic Circuit Depth Reductions with Preserved Adversarial Robustness by Approximate Encoding for Quantum Machine Learning},
   volume={3},
   ISSN={2771-5892},
   url={http://dx.doi.org/10.34133/icomputing.0100},
   DOI={10.34133/icomputing.0100},
   journal={Intelligent Computing},
   publisher={American Association for the Advancement of Science (AAAS)},
   author={West, Maxwell T. and Nakhl, Azar C. and Heredge, Jamie and Creevey, Floyd M. and Hollenberg, Lloyd C. L. and Sevior, Martin and Usman, Muhammad},
   year={2024},
   month=Jan }

@misc{kang2019iot,
    doi = {10.21227/q70p-q449},
    url = {https://dx.doi.org/10.21227/q70p-q449},
    author = {Hyunjae Kang and Dong Hyun Ahn and Gyung Min Lee and Jeong Do Yoo and Kyung Ho Park and Huy Kang Kim},
    publisher = {IEEE Dataport},
    title = {{IoT} network intrusion dataset},
    year = {2019} }

@article{Preskill_2018,
   title={Quantum Computing in the {NISQ} era and beyond},
   volume={2},
   ISSN={2521-327X},
   url={http://dx.doi.org/10.22331/q-2018-08-06-79},
   DOI={10.22331/q-2018-08-06-79},
   journal={Quantum},
   publisher={Verein zur Forderung des Open Access Publizierens in den Quantenwissenschaften},
   author={Preskill, John},
   year={2018},
   month=Aug, pages={79} }

@article{biamonte2017,
   title={Quantum machine learning},
   volume={549},
   ISSN={1476-4687},
   url={http://dx.doi.org/10.1038/nature23474},
   DOI={10.1038/nature23474},
   number={7671},
   journal={Nature},
   publisher={Springer Science and Business Media LLC},
   author={Biamonte, Jacob and Wittek, Peter and Pancotti, Nicola and Rebentrost, Patrick and Wiebe, Nathan and Lloyd, Seth},
   year={2017},
   month=Sep,
   pages={195–202} }

@article{havlicek2019,
   title={Supervised learning with quantum-enhanced feature spaces},
   volume={567},
   ISSN={1476-4687},
   url={http://dx.doi.org/10.1038/s41586-019-0980-2},
   DOI={10.1038/s41586-019-0980-2},
   number={7747},
   journal={Nature},
   publisher={Springer Science and Business Media LLC},
   author={Havlíček, Vojtěch and Córcoles, Antonio D. and Temme, Kristan and Harrow, Aram W. and Kandala, Abhinav and Chow, Jerry M. and Gambetta, Jay M.},
   year={2019},
   month=Mar, pages={209–212} }

@inproceedings{goodfellow2015,
  author    = {Goodfellow, Ian J. and Shlens, Jonathon and Szegedy, Christian},
  title     = {Explaining and Harnessing Adversarial Examples},
  booktitle = {Proc. 3rd Int. Conf. Learning Representations (ICLR)},
  year      = {2015},
  url={https://arxiv.org/abs/1412.6572},
  doi={10.48550/arXiv.1412.6572}
}

@inproceedings{biggio2012,
author = {Biggio, Battista and Nelson, Blaine and Laskov, Pavel},
title = {Poisoning attacks against support vector machines},
year = {2012},
isbn = {9781450312851},
publisher = {Omnipress},
address = {Madison, WI, USA},
booktitle = {Proceedings of the 29th International Conference on International Conference on Machine Learning},
pages = {1467–1474},
numpages = {8},
location = {Edinburgh, Scotland},
series = {ICML'12},
url = {https://dl.acm.org/doi/10.5555/3042573.3042761}
}

@inproceedings{szegedy2014,
  author    = {Szegedy, Christian and Zaremba, Wojciech and Sutskever, Ilya and
               Bruna, Joan and Erhan, Dumitru and Goodfellow, Ian J. and
               Fergus, Rob},
  title     = {Intriguing Properties of Neural Networks},
  booktitle = {Proc. 2nd Int. Conf. Learning Representations (ICLR)},
  year      = {2014},
  url={https://arxiv.org/abs/1312.6199},
  doi={10.48550/arXiv.1312.6199}
}

@article{lu2020,
   title={Quantum adversarial machine learning},
   volume={2},
   ISSN={2643-1564},
   url={http://dx.doi.org/10.1103/PhysRevResearch.2.033212},
   DOI={10.1103/physrevresearch.2.033212},
   number={3},
   journal={Physical Review Research},
   publisher={American Physical Society (APS)},
   author={Lu, Sirui and Duan, Lu-Ming and Deng, Dong-Ling},
   year={2020},
   month=Aug }

@inproceedings{kundu2022,
  author    = {Kundu, Satwik and Ghosh, Swaroop},
  title     = {Adversarial Data Poisoning Attack on Quantum Machine Learning
               in the {NISQ} Era},
  booktitle = {Proc. Great Lakes Symp. VLSI (GLSVLSI)},
  address   = {New Orleans, LA, USA},
  year      = {2025},
  pages     = {976--981},
 url = {https://doi.org/10.1145/3716368.3735244},
 doi = {10.1145/3716368.3735244}
}

@incollection{ghosh2023,
author="Ghosh, Archisman
and Kundu, Satwik
and Ghosh, Swaroop",
editor="Usman, Muhammad",
title="Adversarial Threats in Quantum Machine Learning: A Survey of Attacks and Defenses",
bookTitle="Quantum Robustness in Artificial Intelligence: Principles and Applications",
year="2026",
publisher="Springer Nature Switzerland",
address="Cham",
pages="31--53",
isbn="978-3-032-11153-1",
doi="10.1007/978-3-032-11153-1_3",
url="https://doi.org/10.1007/978-3-032-11153-1_3"
}

@misc{catak2023,
      title={Trustworthy Quantum Machine Learning: A Roadmap for Reliability, Robustness, and Security in the {NISQ} Era}, 
      author={Ferhat Ozgur Catak and Jungwon Seo and Umit Cali},
      year={2025},
      eprint={2511.02602},
      archivePrefix={arXiv},
      primaryClass={quant-ph},
      url={https://arxiv.org/abs/2511.02602}, 
}

@incollection{Powell1994,
author="Powell, M. J. D.",
editor="Gomez, Susana
and Hennart, Jean-Pierre",
title="A Direct Search Optimization Method That Models the Objective and Constraint Functions by Linear Interpolation",
bookTitle="Advances in Optimization and Numerical Analysis",
year="1994",
publisher="Springer Netherlands",
address="Dordrecht",
pages="51--67",
isbn="978-94-015-8330-5",
doi="10.1007/978-94-015-8330-5_4",
url="https://doi.org/10.1007/978-94-015-8330-5_4"
}

@article{abreu2025,
author = {Abreu, Diego and Moura, David and Esteve Rothenberg, Christian and Abelém, Antônio},
title = {{QuantumNetSec}: Quantum Machine Learning for Network Security},
journal = {International Journal of Network Management},
volume = {35},
number = {4},
pages = {e70018},
doi = {https://doi.org/10.1002/nem.70018},
url = {https://onlinelibrary.wiley.com/doi/abs/10.1002/nem.70018},
eprint = {https://onlinelibrary.wiley.com/doi/pdf/10.1002/nem.70018},
year = {2025}
}

@article{kalinin2023,
  title = {Security Intrusion Detection Using Quantum Machine Learning Techniques},
  author = {Kalinin, Maxim and Krundyshev, Vasiliy},
  year = {2023},
  journal = {Journal of Computer Virology and Hacking Techniques},
  shortjournal = {Journal of Computer Virology and Hacking Techniques},
  volume = {19},
  number = {1},
  pages = {125--136},
  issn = {2263-8733},
  doi = {10.1007/s11416-022-00435-0},
  url = {https://doi.org/10.1007/s11416-022-00435-0},
}

@inproceedings{qmlids2023,
   title={{QML-IDS}: Quantum Machine Learning Intrusion Detection System},
   url={http://dx.doi.org/10.1109/ISCC61673.2024.10733655},
   DOI={10.1109/iscc61673.2024.10733655},
   booktitle={2024 IEEE Symposium on Computers and Communications (ISCC)},
   publisher={IEEE},
   author={Abreu, Diego and Rothenberg, Christian Esteve and Abelém, Antonio},
   year={2024},
   month = jun,
   pages={1–6} }

@inproceedings{cicids2017,
author = {Sharafaldin, Iman and Habibi Lashkari, Arash and Ghorbani, Ali},
booktitle = {Proc. 4th Int. Conf. Information Systems Security and Privacy (ICISSP)},
year = {2018},
month = {01},
pages = {108-116},
title = {Toward Generating a New Intrusion Detection Dataset and Intrusion Traffic Characterization},
doi = {10.5220/0006639801080116}
}

@article{wang2021,
   title={Noise-induced barren plateaus in variational quantum algorithms},
   volume={12},
   ISSN={2041-1723},
   url={http://dx.doi.org/10.1038/s41467-021-27045-6},
   DOI={10.1038/s41467-021-27045-6},
   number={1},
   journal={Nature Communications},
   publisher={Springer Science and Business Media LLC},
   author={Wang, Samson and Fontana, Enrico and Cerezo, M. and Sharma, Kunal and Sone, Akira and Cincio, Lukasz and Coles, Patrick J.},
   year={2021},
   month=Nov }

@misc{pennylane,
  title        = {PennyLane: Automatic differentiation of hybrid 
                  quantum-classical computations},
  author       = {Bergholm, Ville and others},
  year         = {2022},
  eprint       = {1811.04968},
  archivePrefix = {arXiv},
  primaryClass = {quant-ph},
   doi={10.48550/arXiv.1811.04968},
  url={https://arxiv.org/abs/1811.04968},
}

@misc{madry2018,
title={Towards deep learning models resistant to adversarial attacks}, 
url={https://hdl.handle.net/1721.1/137496}, 
author={Madry, A and Makelov, A and Schmidt, L and Tsipras, D and Vladu, A}, 
year={2018} }

@inproceedings{marc2026,
  author={Reers, Volker and Maußner, Marc},
  booktitle={2026 International Conference on Quantum Control, Computing and Learning (qCCL)}, 
  title={A Threat-Model–Driven Robustness Benchmark for Quantum Machine Learning Under Device Noise and Deployment Shift}, 
  year={2026},
  volume={},
  number={},
  pages={1-6},
  doi={10.1109/qCCL70331.2026.11668558},
  url = {https://dx.doi.org/10.1109/qCCL70331.2026.11668558}
  }

@inproceedings{maussner2025,
  author    = {Mau{\ss}ner, Marc and Reers, Volker},
  title     = {Study of Robustness of Quantum Machine Learning and Its
               Countermeasures against Adversarial Attacks When Facing Noise},
  booktitle = {INFORMATIK 2025},
  publisher = {Gesellschaft f{\"u}r Informatik e.V.},
  address   = {Bonn},
  year      = {2025},
  pages     = {1751},
  doi       = {10.18420/inf2025_157},
  url = {https://dx.doi.org/10.18420/inf2025_157}
}

@misc{claude2025,
  author       = {{Anthropic}},
  title        = {Claude},
  howpublished = {Large Language Model},
  year         = {2025},
  url          = {https://claude.ai},
}

@misc{ibmquantum,
  author = {{IBM Quantum}},
  title  = {{IBM Quantum}},
  year   = {2025},
  url    = {https://quantum.cloud.ibm.com/}
}

@inproceedings{kingma2017,
  author    = {Kingma, Diederik P. and Ba, Jimmy},
  title     = {Adam: A Method for Stochastic Optimization},
  booktitle = {Proc. 3rd Int. Conf. Learning Representations (ICLR)},
  year      = {2015},
   url={https://arxiv.org/abs/1412.6980},
  doi={10.48550/arXiv.1412.6980}
}

\end{document}